\documentclass[a4paper,fleqn,usenatbib]{mnras}

\usepackage{newtxtext,newtxmath}

\usepackage[T1]{fontenc}
\usepackage{ae,aecompl}

\usepackage{graphicx}	
\usepackage{amsmath}	
\usepackage{amssymb}	

\title[External photoevaporation of IM Lup]{First evidence of external disc photoevaporation in a low mass 
star forming region: the case of IM Lup}
\author[T. J. Haworth et al. ]
{\parbox{\textwidth}{Thomas J. Haworth$^{1}$\thanks{E-mail: \texttt{t.haworth@imperial.ac.uk}}, Stefano Facchini$^{2}$, Cathie J. Clarke$^{3}$ and \\ L. Ilsedore Cleeves$^{4}$\thanks{Hubble fellow}.
}\vspace{0.4cm}\\
\parbox{\textwidth}{$^{1}$Astrophysics Group, Imperial College London, Blackett Laboratory, Prince
Consort Road, London SW7 2AZ, UK \\
$^{2}$Max-Planck-Institut f\"ur Extraterrestrische Physik, Giessenbachstrasse 1, 85748 Garching, Germany \\
$^{3}$Institute of Astronomy, Madingley Rd, Cambridge, CB3 0HA, UK \\
$^{4}$Harvard-Smithsonian Center for Astrophysics, 60 Garden
Street, Cambridge, MA 02138 
}}

\begin{document}
 
\date{Accepted ???. Received ???; in original form ???}

\pagerange{\pageref{firstpage}--\pageref{lastpage}} \pubyear{2016}

\maketitle
\label{firstpage}

\begin{abstract}
We model the radiatively driven flow from IM Lup -- a large protoplanetary disc expected to be irradiated by only a weak external radiation field (at least 10$^4$ times lower than the UV field irradiating the Orion Nebula Cluster proplyds). We find that material at large radii ($>400$\,AU) in this disc is sufficiently weakly gravitationally bound that significant mass loss can be induced. Given the estimated values of the disc mass and accretion rate, the viscous timescale is long ($\sim10$\,Myr) so the main evolutionary behaviour for the first Myr of the disc's lifetime is truncation of the disc by photoevaporation, with only modest changes effected by viscosity.  We also produce approximate synthetic observations of our models, finding substantial emission from the flow which can explain the CO halo observed about IM Lup out to $\geq1000$\,AU.  Solutions that are consistent with the extent of the observed CO emission generally imply that IM Lup is still in the process of having its disc outer radius truncated. We conclude that IM Lup is subject to substantial external photoevaporation, which raises the more general possibility that external irradiation of the largest discs can be of significant importance even in low mass star forming regions.

\end{abstract}


\section{introduction}
Protoplanetary discs - the birthplaces of planets - are found around young stars which are themselves formed in clusters. The discs are thus externally irradiated by other cluster members, in particular by the most massive stars. Strong irradiation of discs close to O stars is well established, for example from observations of proplyds in Orion \citep{1996AJ....111.1977M, 1998ApJ...499..758J, 1998AJ....115..263O, 2000AJ....119.2919B, 2001AJ....122.2662O, 2002ApJ...566..315H}. For some time there has also been the theoretical expectation that protoplanetary discs might be significantly affected by more canonical radiation field strengths found in a cluster environment \citep[e.g.][]{2001MNRAS.325..449S, 2004ApJ...611..360A, 2011PASP..123...14H, 2016MNRAS.457.3593F, 2016MNRAS.tmp.1384H}. This is now being directly supported by recent observations, such as those by \cite{2016ApJ...826L..15K} who identify proplyds irradiatied by a 3000\,G$_0$\footnote{G$_0$ - the Habing unit - is a measure of the UV field local to our solar system and has the value $1.6\times10^{-3}$\,erg\,cm$^{-2}$\,s$^{-1}$.} radiation field -- approximately 100 times weaker than the field strengths irradiating classical proplyds \citep{1999ApJ...515..669S}. To date, however, there is no direct observational inference of externally driven mass loss from discs at lower, but more typical, radiation field strengths in the range $1<$\,G$_0$\,$<1000$. 

IM Lupi is a roughly Solar mass \citep{2009A&A...501..269P} young \citep[$\sim0.5-1$\,Myr;][]{2012A&A...544A.131M} M0 star  situated at a distance of $\sim161$\,pc \citep{2016arXiv160904172G} in the vicinity of the Lupus 2 cloud. Although CO emission likely associated with the disc is detected out to $\sim1000$\,AU, it is only detected in millimetre continuum out to about 313\,AU so the gas disc is more extended than the dust \citep{2007A&A...462..211L, 2008A&A...489..633P, 2009A&A...501..269P, 2016ApJ...832..110C}. The disc is also very massive, with estimates of 0.1 and 0.17\,M$_\odot$ from \cite{2008A&A...489..633P} and \cite{2016ApJ...832..110C} respectively. The mass accretion rate is currently about $10^{-8}$\,M$_\odot$\,yr$^{-1}$ \citep{2016arXiv161207054A}. The most recent analysis of this system by \cite{2016ApJ...832..110C} combined new $^{12}$CO, $^{13}$CO and C$^{18}$O ALMA observations with a broad array of modelling resources to provide a very comprehensive chemical and radiative transfer model of IM Lup, which could describe many features of the disc very successfully. They also included the effect of external irradiation on the composition and thermal structure of the disc. Based on their modelling efforts and from geometrical arguments based on \textit{HIPPARCOS} data they estimate a low UV field incident upon the disc of only  about 4\,G$_0$. There was, however, a diffuse halo of low velocity CO emission about the disc that their model failed to explain. They suggested that this halo might be a remnant structure rather than being material driven out of the disc by photoevaporation. The photoevaporation interpretation was disfavoured based on the inferred low UV field and outer disk temperatures, which were well below those which had been previously considered by external photoevaporation models \citep{2004ApJ...611..360A, 2016MNRAS.457.3593F}. However, since this regime is previously unexplored it is difficult to conclude this with any certainty. 

In this letter we use photochemical-dynamical models to investigate the external irradiation of IM Lup by the weak UV radiation field expected. We aim to determine the expected mass loss rate and flow properties and to determine whether the CO halo could be explained by such a flow. Ultimately we aim to determine whether low radiation field strengths can drive efficient mass loss and whether IM Lup offers an opportunity to observationally probe externally driven mass loss in the modest radiation regime. 

\vspace{-10pt}

\section{Modelling the external photoevaporation of IM Lup}

\subsection{Numerical method and disc construction}
We directly model the photoevaporative outflow, driven by external irradiation, using a radiation hydrodynamics and photodissociation region chemistry code \textsc{torus-3dpdr}, for which key relevant papers are \cite{2012MNRAS.420..562H, 2015MNRAS.448.3156H, 2015MNRAS.453.2277H, 2015MNRAS.454.2828B}. This code was used to run models of externally irradiated discs in benchmark scenarios where there are semi--analytic solutions in \cite{2016MNRAS.tmp.1384H} - validating the approach. The details of the method are also discussed in the latter paper.

In summary we perform calculations of the photodissociation region (PDR) chemistry in sequence with hydrodynamics using operator splitting. The PDR chemistry network is a reduced version of the \textsc{umist} network \citep{2013A&A...550A..36M} including 33 species and 330 reactions, and was derived such that it gives temperatures that do not differ appreciably ($\sim$10\,per cent) from the much more substantial (and computationally expensive) full network. {We do not include polycyclic aromatic hydrocarbons (PAHs), since although they are a key heating mechanism in PDRs, they are observed to be depleted towards discs} \citep{2006A&A...459..545G, 2010ApJ...714..778O}. Our models will therefore yield mass loss rates lower than models that would include PAHs. Because we compute steady state flow profiles we are permitted to perform the PDR calculations relatively infrequently, as the same steady state profile will always eventually result.

Following \cite{2004ApJ...611..360A, 2016MNRAS.457.3593F}, our models are 1D spherical {(see Figure 1 of the latter paper)}. This is believed to be justified because the mass loss is expected to be dominated from the disc outer edge since: i) the material there is least gravitationally bound and ii) the density falls off vertically in a disc more rapidly than radially.  This method also assumes that the incident (exciting) UV field approaches inwards radially and cooling line photons escape outwards radially - so every other direction is infinitely optically thick. 

We employ a fixed structure for the disc itself, which acts as an inner boundary condition to the radiatively driven flow. Interior to some outer disc radius $R_{\textrm{d}}$ we do not allow the conditions to evolve over time. For these fixed disc conditions we use the parameters derived by \cite{2016ApJ...832..110C}. The disc's gas surface density profile follows that of \cite{1974MNRAS.168..603L}, i.e.
\begin{equation}
	\Sigma_{\textrm{g}}(R) = \Sigma_{c}\left(\frac{R}{R_c}\right)^{-\gamma}\exp\left[-\left(\frac{R}{R_c}\right)^{2-\gamma}\right],
	\label{Lynden-BellPringle}
\end{equation}
where \cite{2016ApJ...832..110C} find $\Sigma_c = 25$\,g\,cm$^{-2}$, $R_c =100$\,AU, and $\gamma=1$. 
The scale height is set by
\begin{equation}
	H(R) = H_{100}\left(\frac{R}{100\,\textrm{AU}}\right)^{\psi},
\end{equation}
where \cite{2016ApJ...832..110C} find $\psi=1.15$ and $H_{100}=12$\,AU. 
For the dust we assume a cross section of $\sigma_{\textrm{FUV}} = 5.04\times10^{-23}$\,cm$^{-2}$, dust to gas ratio of $d/g = 10^{-4}$ and the maximum grain size $s_{\textrm{max}}=1\,\mu$m, which are all representative of the kind of dust parameters in the flow found by \cite{2016MNRAS.457.3593F}. We assume that the disc outer edge is sufficiently far from the parent star that the temperature there is only 10\,K. The outer dynamical boundary condition in our models is free-outflow, no-inflow and the inner condition set by the disc properties at $R_{\textrm{d}}$ as described above. The mid-plane number density in the discs of these 1D spherical models is 
\begin{equation}
	n(R) = \frac{1}{\mu m_{\textrm{H}}} \frac{\Sigma_{\textrm{g}}(R)}{\sqrt{2\pi}H(R)}.
	\label{midN}
\end{equation}

The radial extent of our simulation grid -- 10$^{17}$\,cm --  was chosen such that the critical radius in the flow \citep{2016MNRAS.457.3593F} is captured, which we check using the approach detailed in section 5.3.2 of \cite{2016MNRAS.tmp.1384H}. We use an adaptive grid with a maximum number of cells of 2048 and therefore a maximum resolution of 3.25\,AU.  We run each model for 1\,Myr, though steady state flows are established long before this. 

\vspace{-15pt}

\section{Results and discussion}

\subsection{Disc photoevaporation and evolution}
\begin{figure}
	\includegraphics[width=7.7cm]{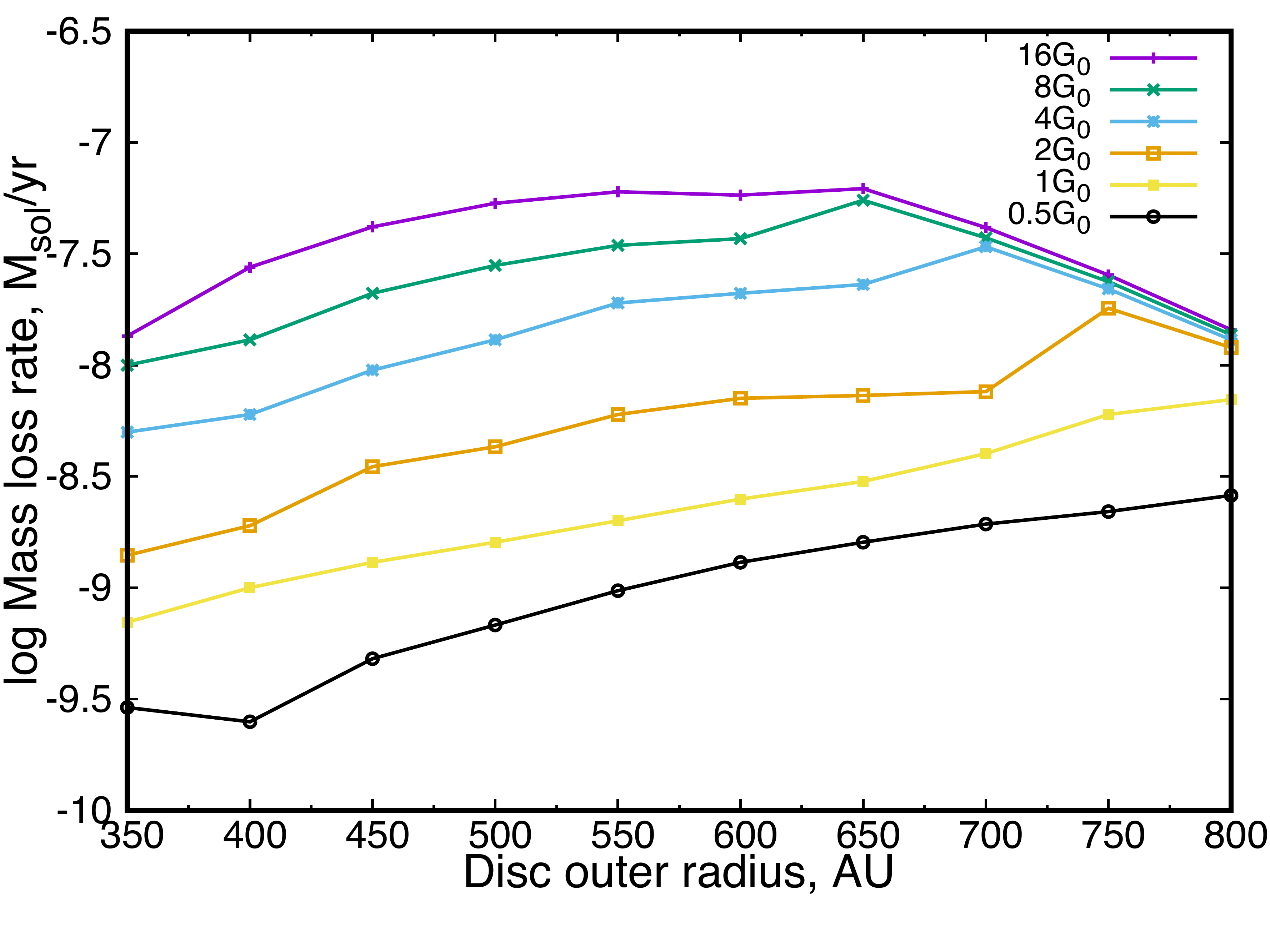}
	\vspace{-10pt}	
	\caption{Log mass loss rate as a function of disc outer radius for different incident UV fields. }
	\vspace{-10pt}
	\label{GridMLR}
\end{figure}

We ran a grid of photoevaporation models for different disc outer radii and incident radiation field strengths. We chose disc radii in 50\,AU intervals from 350 to 800\,AU and radiation field strengths of 0.5, 1, 2, 4, 8 and 16\,G$_0$. We compute the mass loss rate from our models following \cite{2004ApJ...611..360A}:
\begin{equation}
	\dot{M} = 4\pi R^2 \rho \dot{R} \mathcal{F},
\end{equation}
where $\mathcal{F}$ is the fraction of solid angle subtended by the disc outer edge
\begin{equation}
	\mathcal{F} = \frac{H_d}{\sqrt{H_d^2+R_{d}^2}},
\end{equation}
and $H_d$ is the scale height at the disc outer edge $R_d$. 
We compute the average of this quantity over the entire flow (note that $\mathcal{F}$ is constant for a given disc outer radius). A summary of the mass loss rates from our grid of models is shown in Figure \ref{GridMLR}. For large discs (like IM Lup) where material at the outer edge is not so gravitationally bound, substantial mass loss rates ($\sim10^{-8}$\,M$_\odot$\,yr$^{-1}$) can be driven even when the incident radiation field strength is very modest. Note that the 4\,G$_0$ field expected to be irradiating IM Lup and driving this mass loss is $\sim10^3$ times weaker than that irradiating the proplyds observed by \cite{2016ApJ...826L..15K}  and $\sim10^4$ times weaker than the proplyds in the ONC \citep[e.g.][]{2000AJ....119.2919B, 2001AJ....122.2662O, 2002ApJ...566..315H}

 The current mass accretion rate in this system was recently computed using new X--shooter data to be $10^{-8}$\,M$_{\odot}$\,yr$^{-1}$ with an uncertainty of 0.35 dex by \cite{2016arXiv161207054A}. The external photoevaporative mass loss rate for UV fields $\geq4$\,G$_0$ is hence expected to be of order or greater than the mass accretion rate. 

\begin{figure}
	\includegraphics[width=7.7cm]{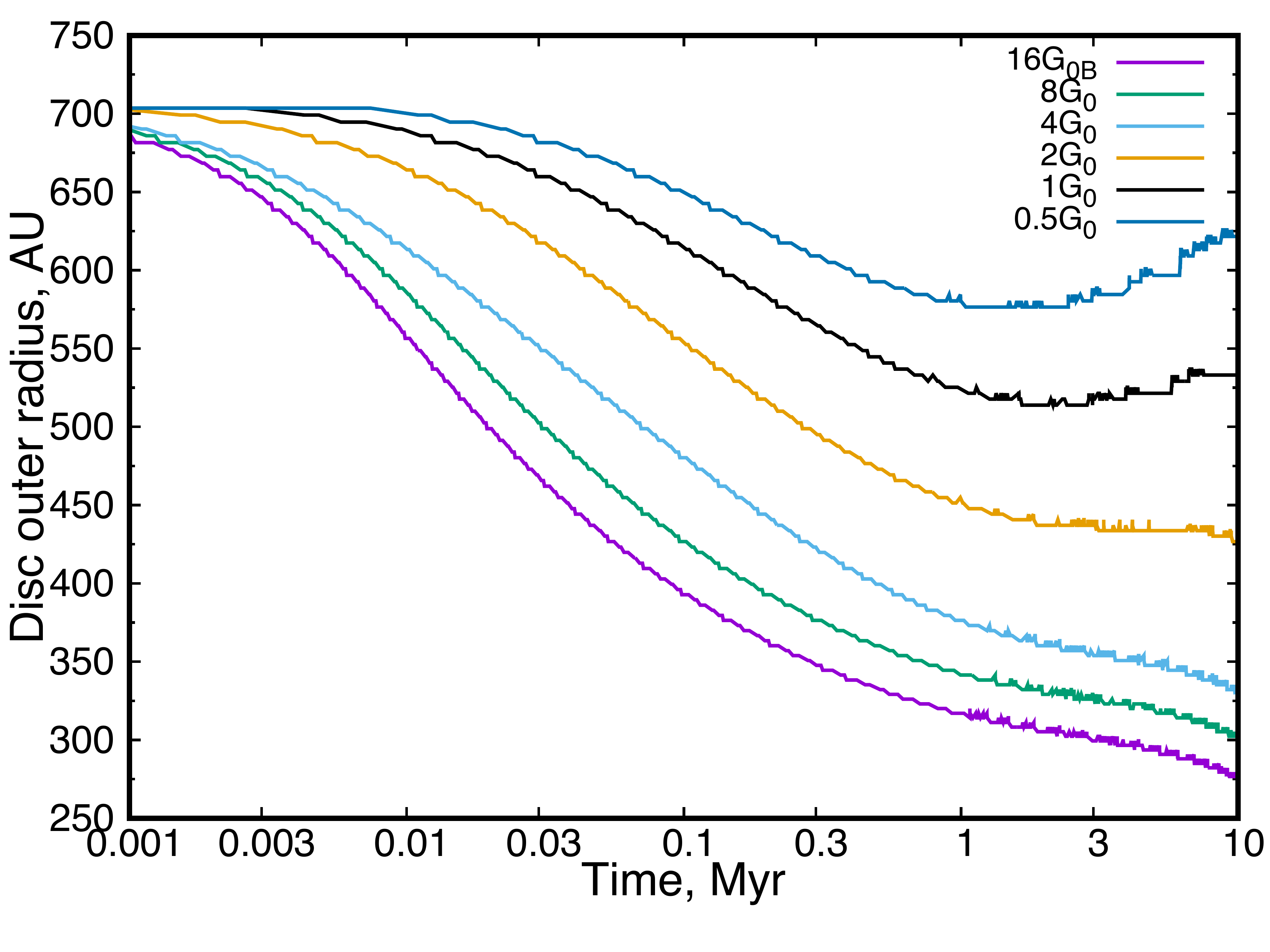}
	\vspace{-10pt}		
	\caption{The outer disc radius of IM Lup  as a function of time according to our evolutionary models that include external photoevaporation.}
	\vspace{-10pt}
	
	\label{Router}
\end{figure}

\begin{figure}
	\includegraphics[width=7.7cm]{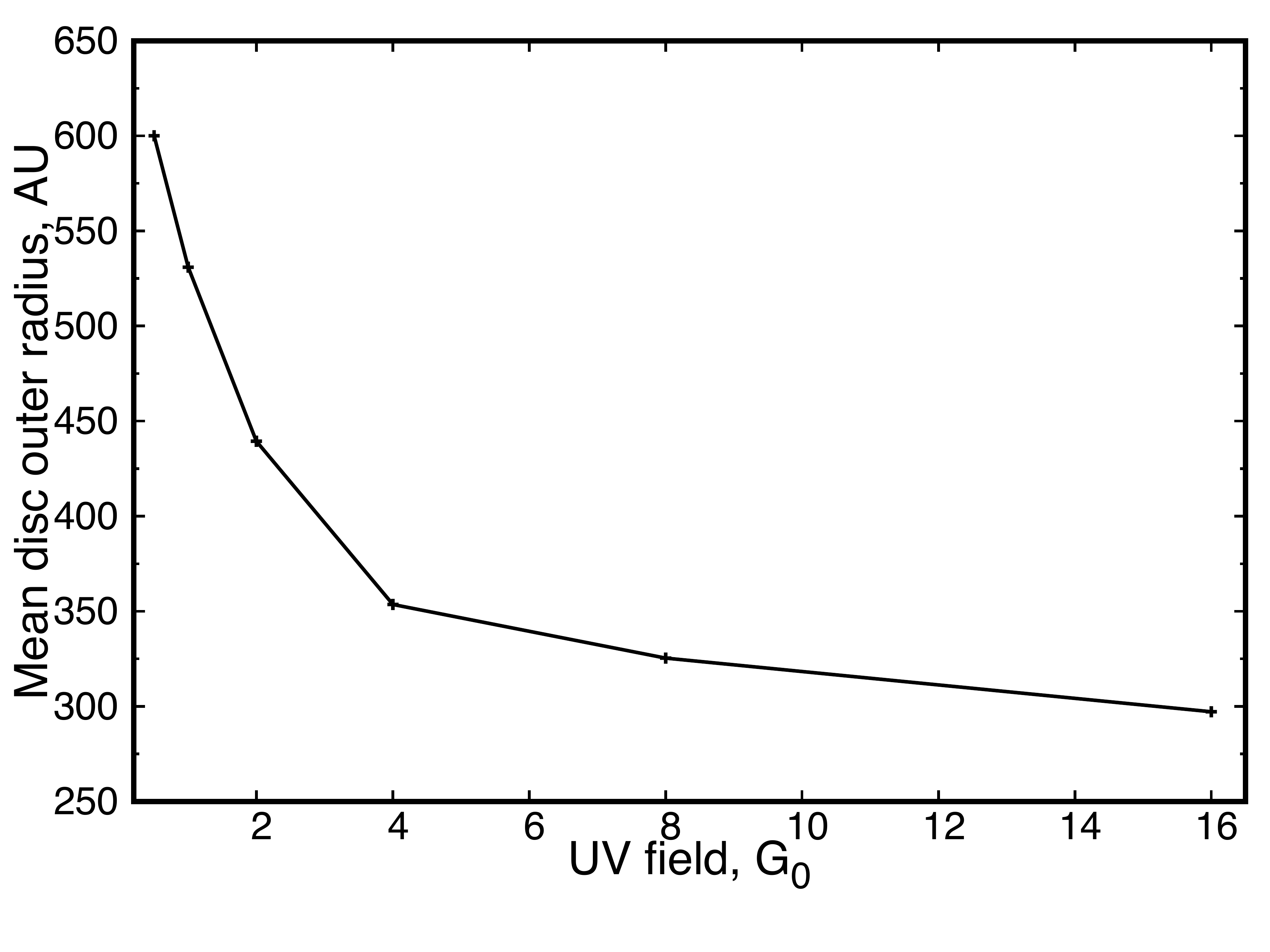}
	\vspace{-10pt}
	\caption{{The mean radius of IM Lup over 10\,Myr as a function of the incident UV field strength. }}
	\vspace{-10pt}
	\label{Rstag}
\end{figure}

We fit the mass loss rate as a function of radius, which we feed into the \cite{2007MNRAS.376.1350C} secular evolutionary code to examine the disc evolution. The viscous timescale of this disc is of order 10\,Myr, so the main evolutionary behaviour is truncation of the disc by external photoevaporation.  Figure \ref{Router} shows the evolution of the disc outer edge as a function of time for different incident UV field strengths {(note that for models that drop below an outer radius of 350\,AU we compute additional photoevaporation models to estimate the mass loss rate at these smaller radii)}. In all cases the disc outer edge {rapidly retreats to some stagnation radius in less than 1\,Myr, after which it varies in size only slowly}. The {mean radius over 10\,Myr} as a function of incident UV field is given in Figure \ref{Rstag}, showing strong variation for fields $<8$\,G$_0$. A key point is that because the observed CO emission is currently extended out to beyond 1000\,AU, even an extremely weak UV radiation field would be expected to truncate this very rapidly. The observed CO emission therefore either has to be part of a photoevaporative flow, or part of some \textit{much} denser envelope that is resilient against the effects of the incident radiation field. Because IM Lup is very young ($\leq 1$\,Myr), its outer edge may still be in the process of retreating towards the stagnation radius.  Another interesting point is that due to the disc's long viscous timescale IM Lup is likely to remain unusually large at the stagnation radius (perhaps > 300\,AU) for many Myr, unless some other mechanism further truncates the disc.

\vspace{-10pt}

\subsection{Could external photoevaporation explain the CO halo?}
Our models imply that even in the presence of a weak UV field substantial mass loss is being induced from IM Lup by external photoevaporation; however, being 1D, they are difficult to directly compare with the real observed data of \cite{2016ApJ...832..110C}. Generating synthetic observations from 1D models has the limitation that some assumption about the vertical density, temperature and compositional structure is required. Nevertheless we make an optimistic attempt at comparison.  We assume that the disc (the boundary condition of the dynamical models) is hydrostatic.  In the flow region we use our simulation results and assume that at a given spherical radius there is a constant density, isothermal, isochemical flow, with scale height set by assuming that $H/R$ beyond the disc outer edge is constant.  We produce synthetic data cubes using the comoving frame molecular line radiative transfer components of \textsc{torus}, detailed in \cite{2010MNRAS.407..986R}. These cubes are then azimuthally averaged in the same manner used to produce the results in Figure 12 of \cite{2016ApJ...832..110C}. 

Because our synthetic observations are based on 1D models, and there is a large array of possible parameters, we do not aim to fit the CO observations. Furthermore, given that we are comparing with $^{12}$CO our synthetic observations will be particularly limited in components of the flow that are optically thick (which vary for each model but we generally find are interior to about 800--900\,AU).  Rather then, we aim to demonstrate that even weak external photoevaporation is capable of producing substantial emission at large radii, such as that observed in the CO halo of IM Lup.

\begin{figure}
	\includegraphics[width=7.8cm]{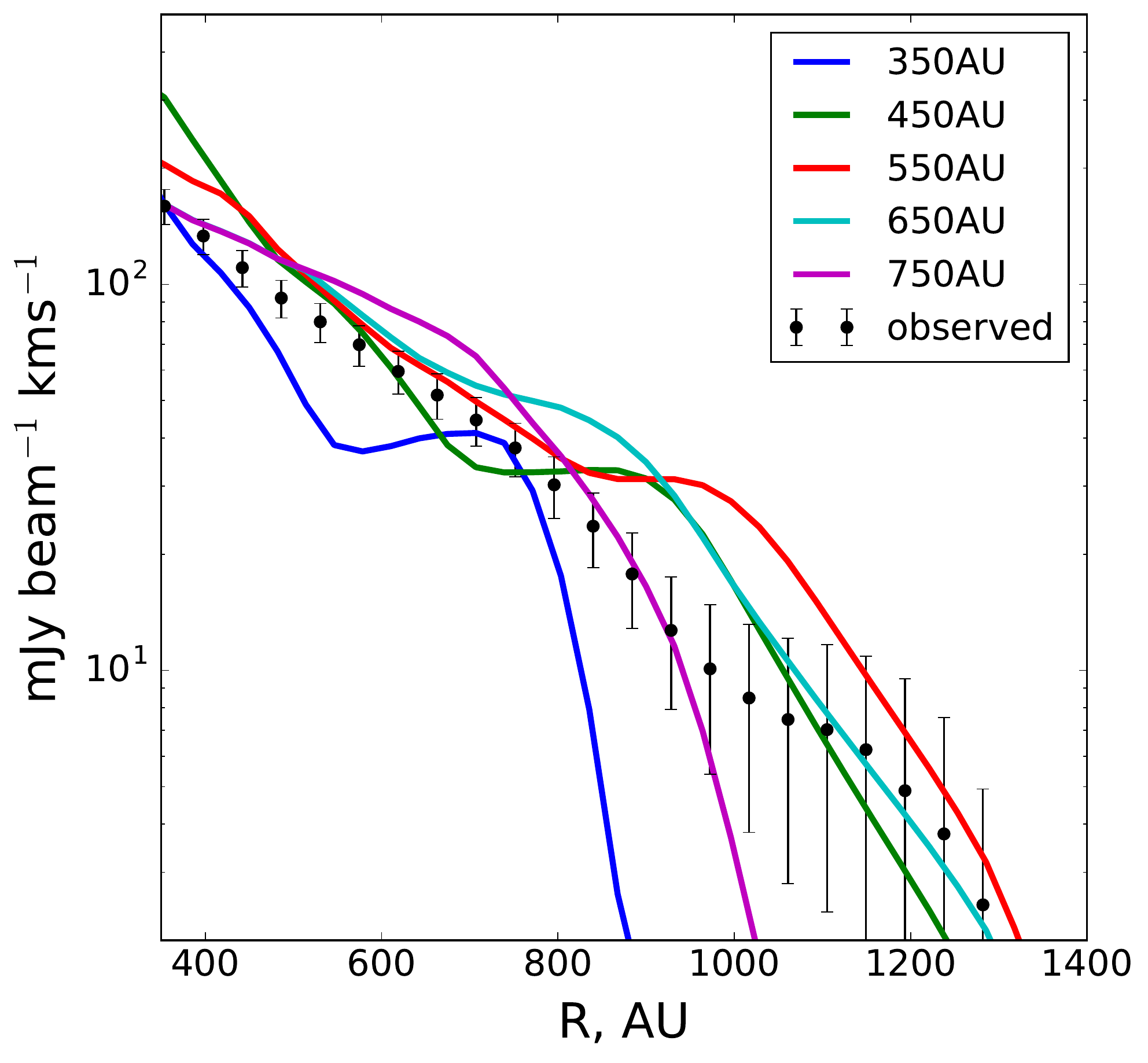}
	\includegraphics[width=7.8cm]{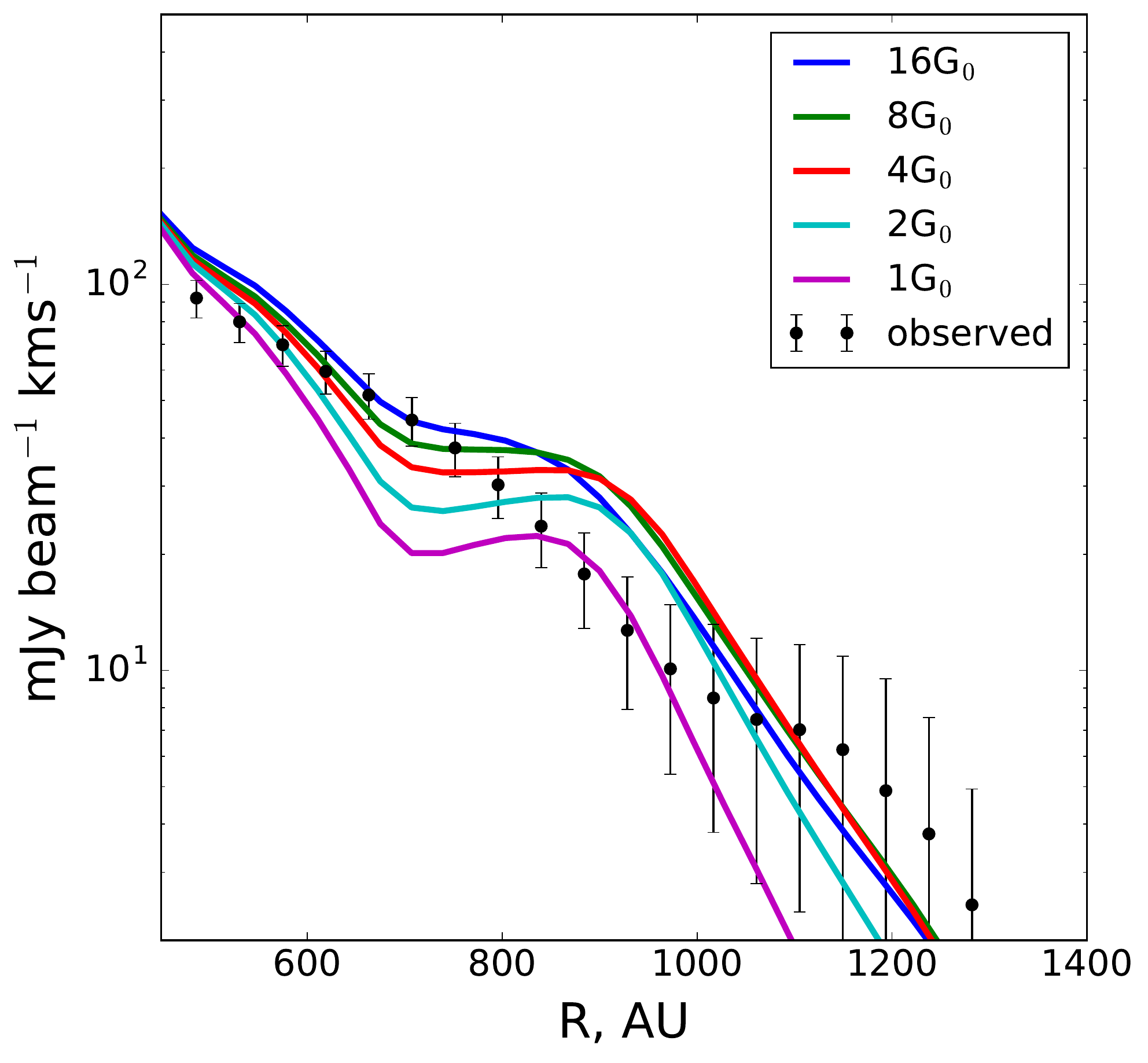}
	\includegraphics[width=7.8cm]{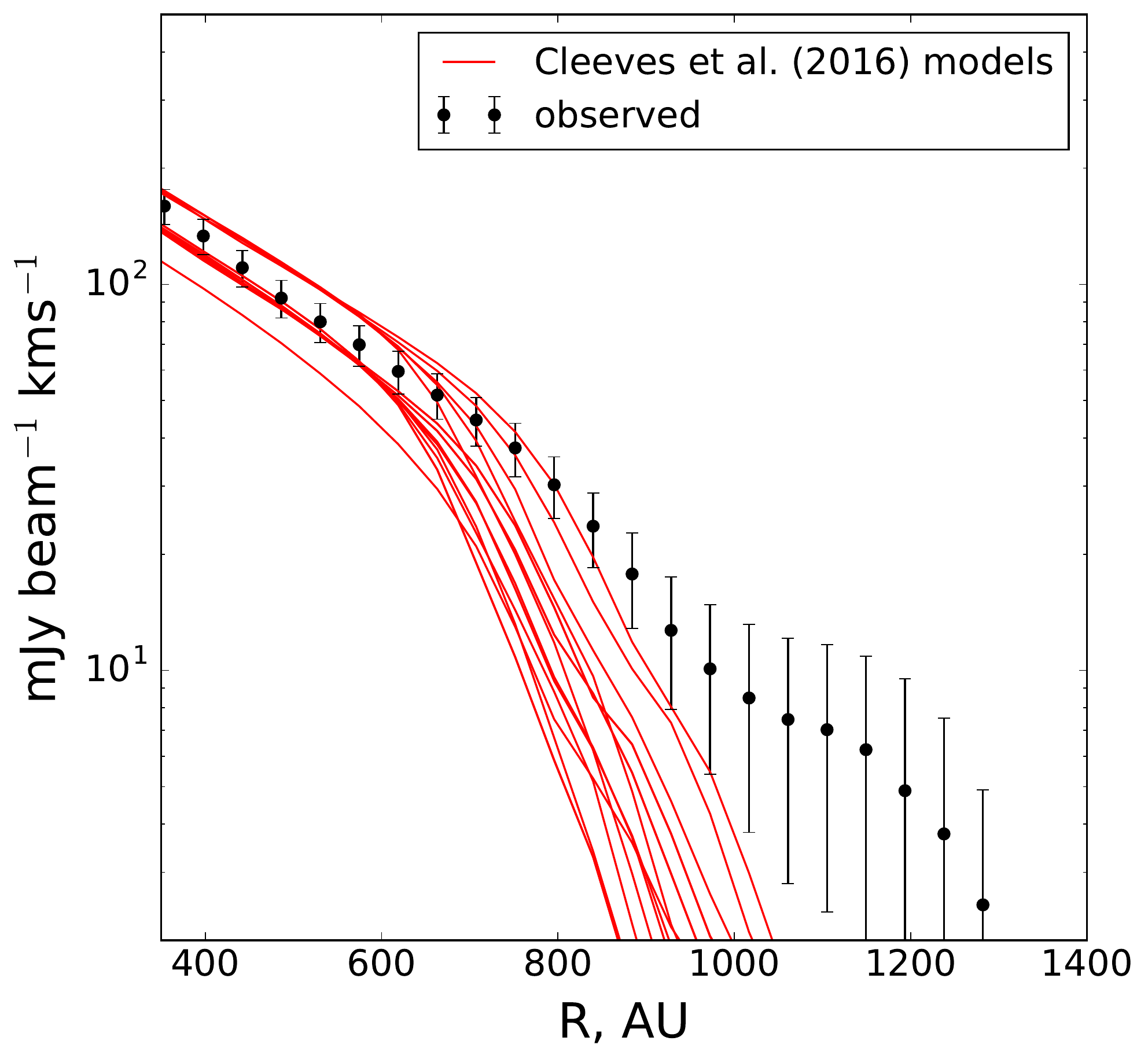}
	\vspace{-10pt}
	\caption{Azimuthally averaged emission profiles from our approximate synthetic observations, compared to the observed data points (with 1\,$\sigma$ error bars) from Cleeves et al. (2016). The upper panel varies the disc outer radius for a radiation field of 4\,G$_0$. The middle panel varies the incident radiation field strength upon a {450\,AU disc}.  The bottom panel shows a collection of models from Cleeves et al. (2016). }
	\label{synthObservations}
\end{figure}

Figure \ref{synthObservations} shows a collection of approximate synthetic emission profiles from our photoevaporation models, as well as emission profiles from a selection of the models from \cite{2016ApJ...832..110C}. The latter models modify the incident UV field but do not permit radial dynamical evolution and thus impose the surface density profile given by equation \ref{Lynden-BellPringle}. As a result the extent of the CO emission is significantly less than that observed. Conversely our external photoevaporation models do show emission comparable in extent and magnitude to the observations.

\begin{figure}
	\hspace{-10pt}
	\vspace{-15pt}
	
	\includegraphics[width=8cm]{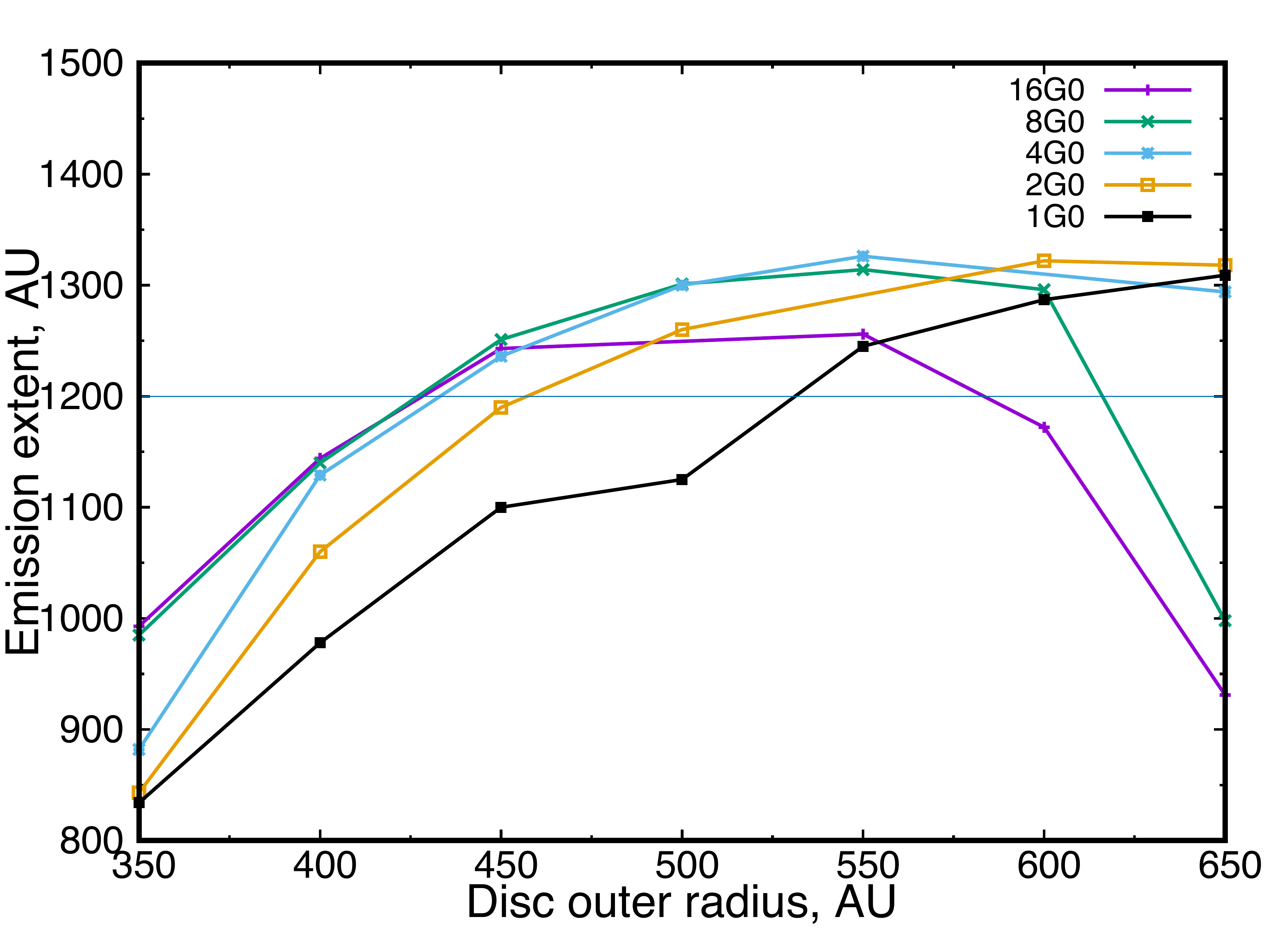}
\vspace{-10pt}
	\caption{The approximate radial extent of CO emission in our models as a function of disc outer radius. The horizontal line represents the radial extent from the best fit model of Cleeves et al. (2016).  }
\vspace{-10pt}
	\label{extent}
\end{figure}

In Figure \ref{extent} we plot the radial extent of the CO emission in our models as a function of the disc outer radius, with different lines representing different incident UV fluxes. The fitted gas extent from \citet{2016ApJ...832..110C} is 1200\,AU which we take to be ``the extent'' of IM Lup for our comparison here, though in practice the detection is marginal beyond 1000\,AU. From our models the extent is the point at which the flux drops below $2$\,mJy\,beam$^{-1}$\,km\,s$^{-1}$, which is the background as calculated using the average of the first and last velocity channels in the synthetic data cube. Most of our models have an extent $1000-1300$\,AU. Generally the models that have extent consistent with the observations have disc outer radii which imply that the disc outer edge is still retreating. 

If the observed extent were known with higher certainty, we could use it in conjunction with Figures \ref{Router} and \ref{extent}  to constrain the minimum disc outer edge and hence maximum age. For example if we knew that the observed extent was 1200\,AU then linearly interpolating Figure \ref{extent} would yield minimum disc outer radii of {430, 450 and 530\,AU} for incident UV fields of 4, 2 and 1\,G$_0$ respectively. Using our evolutionary models from Figure \ref{Router}, these minimum disc outer radii would correspond to approximate maximum IM Lup ages of {0.3, 0.8 and 0.8} \,Myr respectively -- so all would be conceivable given the uncertain 0.5--1\,Myr estimate for the age of IM Lup. Future higher sensitivity observations might offer such a constraint.


\vspace{-15pt}

\section{Summary and conclusions}
We model the external photoevaporation of the large protoplanetary disc IM Lup. This disc has a large CO ``halo'' that was identified in recent ALMA observations by \cite{2016ApJ...832..110C} which could not be explained by hydrostatic chemical and radiative transfer models which assumed that the surface density at large radius was an extrapolation of the form given in equation \ref{Lynden-BellPringle}. We find that although the radiation field irradiating IM Lup is very weak ($<10^4$ times the UV field irradiating the proplyds near O stars in Orion), the disc is sufficiently large that the weakly gravitationally bound material at the disc outer edge can be efficiently photoevaporated. Specifically a 4\,G$_0$ radiation field induces mass loss of  $\sim10^{-8}$\,M$_{\odot}$\,yr$^{-1}$ which is comparable to the current accretion rate onto the star. 
Having a $\sim$10\,Myr viscous timescale, the effect of this mass loss is to rapidly (<1\,Myr) truncate the disc outer edge down to some stagnation radius. The stagnation radius ranges from about 600\,AU for an irradiating UV field of 0.5\,G$_0$ down to about {300-350\,AU} for fields $8-16$\,G$_0$. In the absence of other external influences the disc {only evolves slowly away from the stagnation radius over 10\,Myr}. {Once gas from the disc cannot be delivered to the outer edge at a rate sufficient to supply the photoevaporative wind the disc is expected to shrink rapidly}. 

We also generated approximate synthetic observations from our models, which are able to explain the radial extent of CO emission about IM Lup. Our scenarios that are consistent with the observed extent of CO emission of IM Lup generally imply that its disc outer radius is still in the process of being truncated.  More generally we demonstrate that even weak external fields can lead to significant extended emission from large discs, which hydrostatic models are unable to achieve. 


\vspace{-15pt}

\section*{Acknowledgements}
TJH is funded by an Imperial College London Junior Research Fellowship. This work has also been supported by the DISCSIM project, grant agreement 341137 funded by the European Research Council under ERC-2013-ADG. LIC acknowledges the support of NASA through Hubble Fellowship grant HST-HF2-51356.001-A awarded by the Space Telescope Science Institute, which is operated by the Association of Universities for Research in Astronomy, Inc., for NASA, under contract NAS 5-26555.
This work was undertaken on the COSMOS Shared Memory system at DAMTP, University of Cambridge operated on behalf of the STFC DiRAC HPC Facility. This equipment is funded by BIS National E-infrastructure capital grant ST/J005673/1 and STFC grants ST/H008586/1, ST/K00333X/1. 
This paper makes use of the following ALMA data: ADS/JAO.ALMA\#2013.00694. ALMA is a partnership of ESO (representing its member states), NSF (USA) and NINS (Japan), together with NRC (Canada) and NSC and ASIAA (Taiwan), in cooperation with the Republic of Chile. The Joint ALMA Observatory is operated by ESO, AUI/NRAO and NAOJ. The National Radio Astronomy Observatory is a facility of the National Science Foundation operated under cooperative agreement by Associated Universities, Inc. 

\vspace{-18pt}

\bibliographystyle{mn2e}
\bibliography{molecular}

\begin{thebibliography}{}

\bibitem[\protect\citeauthoryear{{Adams}, {Hollenbach}, {Laughlin} \&
  {Gorti}}{{Adams} et~al.}{2004}]{2004ApJ...611..360A}
{Adams} F.~C.,  {Hollenbach} D.,  {Laughlin} G.,    {Gorti} U.,  2004, \apj,
  611, 360

\bibitem[\protect\citeauthoryear{{Alcala'}, {Manara}, {Natta}, {Frasca},
  {Testi}, {Nisini}, {Stelzer}, {Williams}, {Antoniucci}, {BIazzo}, {Covino},
  {Esposito}, {Getman} \& {Rigliaco}}{{Alcala'}
  et~al.}{2016}]{2016arXiv161207054A}
{Alcala'} J.~M.,  {Manara} C.~F.,  {Natta} A.,  {Frasca} A.,  {Testi} L.,
  {Nisini} B.,  {Stelzer} B.,  {Williams} J.~P.,  {Antoniucci} S.,  {BIazzo}
  K.,  {Covino} E.,  {Esposito} M.,  {Getman} F.,    {Rigliaco} E.,  2016,
  ArXiv e-prints

\bibitem[\protect\citeauthoryear{{Bally}, {O'Dell} \& {McCaughrean}}{{Bally}
  et~al.}{2000}]{2000AJ....119.2919B}
{Bally} J.,  {O'Dell} C.~R.,    {McCaughrean} M.~J.,  2000, \aj, 119, 2919

\bibitem[\protect\citeauthoryear{{Bisbas}, {Haworth}, {Barlow}, {Viti},
  {Harries}, {Bell} \& {Yates}}{{Bisbas} et~al.}{2015}]{2015MNRAS.454.2828B}
{Bisbas} T.~G.,  {Haworth} T.~J.,  {Barlow} M.~J.,  {Viti} S.,  {Harries}
  T.~J.,  {Bell} T.,    {Yates} J.~A.,  2015, \mnras, 454, 2828

\bibitem[\protect\citeauthoryear{{Clarke}}{{Clarke}}{2007}]{2007MNRAS.376.1350C}
{Clarke} C.~J.,  2007, \mnras, 376, 1350

\bibitem[\protect\citeauthoryear{{Cleeves}, {{\"O}berg}, {Wilner}, {Huang},
  {Loomis}, {Andrews} \& {Czekala}}{{Cleeves}
  et~al.}{2016}]{2016ApJ...832..110C}
{Cleeves} L.~I.,  {{\"O}berg} K.~I.,  {Wilner} D.~J.,  {Huang} J.,  {Loomis}
  R.~A.,  {Andrews} S.~M.,    {Czekala} I.,  2016, \apj, 832, 110

\bibitem[\protect\citeauthoryear{{Facchini}, {Clarke} \& {Bisbas}}{{Facchini}
  et~al.}{2016}]{2016MNRAS.457.3593F}
{Facchini} S.,  {Clarke} C.~J.,    {Bisbas} T.~G.,  2016, \mnras, 457, 3593

\bibitem[\protect\citeauthoryear{{Gaia Collaboration}, {Brown}, {Vallenari},
  {Prusti}, {de Bruijne}, {Mignard}, {Drimmel} \& {co-authors}}{{Gaia
  Collaboration} et~al.}{2016}]{2016arXiv160904172G}
{Gaia Collaboration} {Brown} A.~G.~A.,  {Vallenari} A.,  {Prusti} T.,  {de
  Bruijne} J.,  {Mignard} F.,  {Drimmel} R.,    {co-authors} .,  2016, ArXiv
  e-prints

\bibitem[\protect\citeauthoryear{{Geers}, {Augereau}, {Pontoppidan},
  {Dullemond}, {Visser}, {Kessler-Silacci}, {Evans} II, {van Dishoeck},
  {Blake}, {Boogert}, {Brown}, {Lahuis} \& {Mer{\'{\i}}n}}{{Geers}
  et~al.}{2006}]{2006A&A...459..545G}
{Geers} V.~C.,  {Augereau} J.-C.,  {Pontoppidan} K.~M.,  {Dullemond} C.~P.,
  {Visser} R.,  {Kessler-Silacci} J.~E.,  {Evans} II N.~J.,  {van Dishoeck}
  E.~F.,  {Blake} G.~A.,  {Boogert} A.~C.~A.,  {Brown} J.~M.,  {Lahuis} F.,
  {Mer{\'{\i}}n} B.,  2006, \aap, 459, 545

\bibitem[\protect\citeauthoryear{{Harries}}{{Harries}}{2015}]{2015MNRAS.448.3156H}
{Harries} T.~J.,  2015, \mnras, 448, 3156

\bibitem[\protect\citeauthoryear{{Haworth}, {Boubert}, {Facchini}, {Bisbas} \&
  {Clarke}}{{Haworth} et~al.}{2016}]{2016MNRAS.tmp.1384H}
{Haworth} T.~J.,  {Boubert} D.,  {Facchini} S.,  {Bisbas} T.~G.,    {Clarke}
  C.~J.,  2016, \mnras

\bibitem[\protect\citeauthoryear{{Haworth} \& {Harries}}{{Haworth} \&
  {Harries}}{2012}]{2012MNRAS.420..562H}
{Haworth} T.~J.,  {Harries} T.~J.,  2012, \mnras, 420, 562

\bibitem[\protect\citeauthoryear{{Haworth}, {Harries}, {Acreman} \&
  {Bisbas}}{{Haworth} et~al.}{2015}]{2015MNRAS.453.2277H}
{Haworth} T.~J.,  {Harries} T.~J.,  {Acreman} D.~M.,    {Bisbas} T.~G.,  2015,
  \mnras, 453, 2277

\bibitem[\protect\citeauthoryear{{Henney}, {O'Dell}, {Meaburn}, {Garrington} \&
  {Lopez}}{{Henney} et~al.}{2002}]{2002ApJ...566..315H}
{Henney} W.~J.,  {O'Dell} C.~R.,  {Meaburn} J.,  {Garrington} S.~T.,    {Lopez}
  J.~A.,  2002, \apj, 566, 315

\bibitem[\protect\citeauthoryear{{Holden}, {Landis}, {Spitzig} \&
  {Adams}}{{Holden} et~al.}{2011}]{2011PASP..123...14H}
{Holden} L.,  {Landis} E.,  {Spitzig} J.,    {Adams} F.~C.,  2011, \pasp, 123,
  14

\bibitem[\protect\citeauthoryear{{Johnstone}, {Hollenbach} \&
  {Bally}}{{Johnstone} et~al.}{1998}]{1998ApJ...499..758J}
{Johnstone} D.,  {Hollenbach} D.,    {Bally} J.,  1998, \apj, 499, 758

\bibitem[\protect\citeauthoryear{{Kim}, {Clarke}, {Fang} \& {Facchini}}{{Kim}
  et~al.}{2016}]{2016ApJ...826L..15K}
{Kim} J.~S.,  {Clarke} C.~J.,  {Fang} M.,    {Facchini} S.,  2016, \apjl, 826,
  L15

\bibitem[\protect\citeauthoryear{{Lommen}, {Wright}, {Maddison},
  {J{\o}rgensen}, {Bourke}, {van Dishoeck}, {Hughes}, {Wilner}, {Burton} \&
  {van Langevelde}}{{Lommen} et~al.}{2007}]{2007A&A...462..211L}
{Lommen} D.,  {Wright} C.~M.,  {Maddison} S.~T.,  {J{\o}rgensen} J.~K.,
  {Bourke} T.~L.,  {van Dishoeck} E.~F.,  {Hughes} A.,  {Wilner} D.~J.,
  {Burton} M.,    {van Langevelde} H.~J.,  2007, \aap, 462, 211

\bibitem[\protect\citeauthoryear{{Lynden-Bell} \& {Pringle}}{{Lynden-Bell} \&
  {Pringle}}{1974}]{1974MNRAS.168..603L}
{Lynden-Bell} D.,  {Pringle} J.~E.,  1974, \mnras, 168, 603

\bibitem[\protect\citeauthoryear{{Mawet}, {Absil}, {Montagnier}, {Riaud},
  {Surdej}, {Ducourant}, {Augereau}, {R{\"o}ttinger}, {Girard}, {Krist} \&
  {Stapelfeldt}}{{Mawet} et~al.}{2012}]{2012A&A...544A.131M}
{Mawet} D.,  {Absil} O.,  {Montagnier} G.,  {Riaud} P.,  {Surdej} J.,
  {Ducourant} C.,  {Augereau} J.-C.,  {R{\"o}ttinger} S.,  {Girard} J.,
  {Krist} J.,    {Stapelfeldt} K.,  2012, \aap, 544, A131

\bibitem[\protect\citeauthoryear{{McCaughrean} \& {O'dell}}{{McCaughrean} \&
  {O'dell}}{1996}]{1996AJ....111.1977M}
{McCaughrean} M.~J.,  {O'dell} C.~R.,  1996, \aj, 111, 1977

\bibitem[\protect\citeauthoryear{{McElroy}, {Walsh}, {Markwick}, {Cordiner},
  {Smith} \& {Millar}}{{McElroy} et~al.}{2013}]{2013A&A...550A..36M}
{McElroy} D.,  {Walsh} C.,  {Markwick} A.~J.,  {Cordiner} M.~A.,  {Smith} K.,
   {Millar} T.~J.,  2013, \aap, 550, A36

\bibitem[\protect\citeauthoryear{{O'Dell}}{{O'Dell}}{1998}]{1998AJ....115..263O}
{O'Dell} C.~R.,  1998, \aj, 115, 263

\bibitem[\protect\citeauthoryear{{O'Dell}}{{O'Dell}}{2001}]{2001AJ....122.2662O}
{O'Dell} C.~R.,  2001, \aj, 122, 2662

\bibitem[\protect\citeauthoryear{{Oliveira}, {Pontoppidan}, {Mer{\'{\i}}n},
  {van Dishoeck}, {Lahuis}, {Geers}, {J{\o}rgensen}, {Olofsson}, {Augereau} \&
  {Brown}}{{Oliveira} et~al.}{2010}]{2010ApJ...714..778O}
{Oliveira} I.,  {Pontoppidan} K.~M.,  {Mer{\'{\i}}n} B.,  {van Dishoeck} E.~F.,
   {Lahuis} F.,  {Geers} V.~C.,  {J{\o}rgensen} J.~K.,  {Olofsson} J.,
  {Augereau} J.-C.,    {Brown} J.~M.,  2010, \apj, 714, 778

\bibitem[\protect\citeauthoryear{{Pani{\'c}}, {Hogerheijde}, {Wilner} \&
  {Qi}}{{Pani{\'c}} et~al.}{2009}]{2009A&A...501..269P}
{Pani{\'c}} O.,  {Hogerheijde} M.~R.,  {Wilner} D.,    {Qi} C.,  2009, \aap,
  501, 269

\bibitem[\protect\citeauthoryear{{Pinte}, {Padgett}, {M{\'e}nard},
  {Stapelfeldt}, {Schneider}, {Olofsson}, {Pani{\'c}}, {Augereau},
  {Duch{\^e}ne} \& {et al.}}{{Pinte} et~al.}{2008}]{2008A&A...489..633P}
{Pinte} C.,  {Padgett} D.~L.,  {M{\'e}nard} F.,  {Stapelfeldt} K.~R.,
  {Schneider} G.,  {Olofsson} J.,  {Pani{\'c}} O.,  {Augereau} J.~C.,
  {Duch{\^e}ne} G.,    {et al.} 2008, \aap, 489, 633

\bibitem[\protect\citeauthoryear{{Rundle}, {Harries}, {Acreman} \&
  {Bate}}{{Rundle} et~al.}{2010}]{2010MNRAS.407..986R}
{Rundle} D.,  {Harries} T.~J.,  {Acreman} D.~M.,    {Bate} M.~R.,  2010,
  \mnras, 407, 986

\bibitem[\protect\citeauthoryear{{Scally} \& {Clarke}}{{Scally} \&
  {Clarke}}{2001}]{2001MNRAS.325..449S}
{Scally} A.,  {Clarke} C.,  2001, \mnras, 325, 449

\bibitem[\protect\citeauthoryear{{St{\"o}rzer} \& {Hollenbach}}{{St{\"o}rzer}
  \& {Hollenbach}}{1999}]{1999ApJ...515..669S}
{St{\"o}rzer} H.,  {Hollenbach} D.,  1999, \apj, 515, 669

\end{thebibliography}

\label{lastpage}

\end{document}